\documentclass{iau} 
\usepackage{graphicx}

\title[Gamma-ray emission in radio galaxies, from MeV to TeV] 
{Gamma-ray emission in radio galaxies, from MeV to TeV}

\author[Eleonora Torresi]   
{Eleonora Torresi$^{1,2}$ 
 }

\affiliation{On behalf of the Fermi-LAT Collaboration \\
$^1$Dipartimento di Astronomia, Universit\`a di Bologna, 
via Gobetti, 93/2, I-40129, Bologna, Italy \\ 
$^2$INAF-OAS Bologna, Area della Ricerca CNR, via Gobetti, 101, I-40129, Bologna, Italy \\
email: {\tt eleonora.torresi@inaf.it} \\[\affilskip]}

\pubyear{2018}
\volume{342}  
\jname{Perseus in Sicily: from black holes to cluster outskirts}
\editors{Asada, K., de Gouveia dal Pino, E., Nagai, H., \\
Nemmen, R. \& Giroletti, M. eds.}
\begin{document}

\maketitle

\begin{abstract}
Thanks to the \emph{Fermi} $\gamma$-ray satellite and the current Imaging Atmospheric Cherenkov Telescopes, radio galaxies have arisen as a new class of high- and very-high energy emitters. The favourable orientation of their jets makes radio galaxies extremely relevant in addressing important issues such as: (i) revealing the jet structure complexity; (ii) localising the emitting region(s) of high- and very-high energy radiation; (iii) understanding the physical processes producing these photons. In this review the main results on the $\gamma$-ray emission studies of radio galaxies from the MeV to TeV regimes will be presented, and the impact of future Cherenkov Telescope Array observations will be discussed.

\end{abstract}

\firstsection 
\section{Introduction}
\noindent
Radio galaxies (RG) are radio-loud Active Galactic Nuclei (AGN) characterised by powerful jets of relativistic plasma whose non-thermal radiation emits from radio frequencies up to TeV energies.
RGs have been historically classified by \cite[Fanaroff \& Riley (1974)]{fr74} on the basis of their radio morphology into  FR~I and FR~II \footnote{A source is considered an FR~I if the separation between the points of the peak intensity in the two lobes is smaller than half of the largest size of the source (R$<$0.5). If R$>$0.5 the source is a FR~II. This morphological criterium corresponds to a separation at a radio luminosity P$_{178~MHz}$=10$^{25}$~W~Hz$^{-1}$~sr$^{-1}$. Sources with a radio luminosity below such a threshold are FR~Is, above it are FR~IIs.}. Within the Unified Model of AGN (\cite[Urry \& Padovani 1995]{UrryPadovani95}), FR~Is are considered the parent population of BL Lacs (i.e. their misaligned counterpart) and FR~IIs the parent population of Flat Spectrum Radio Quasars (FSRQ). The cause of the FRI/FRII dichotomy is still unknown: external agents, e.g. environment, host galaxy, merging history, etc. (\cite[Bicknell 1995]{bicknell95}) or intrinsic factors, e.g. accretion processes (\cite[Ghisellini \& Celotti 2001]{GhiselliniCelotti01}), have been invoked as possible explanations.
Jets of RGs are oriented at large inclination angles with respect to the l.o.s.  ($\theta >$10$^{\circ}$), making these sources unique laboratories where to contemporarily study jets and accretion processes and establish a connection between the two.
However, the large jet inclination angle geometrically disfavours the detection of RGs above 100~MeV since their non-thermal radiation does not benefit from the strong Doppler amplification typical of blazars. Indeed, the degree of boosting depends on the relativistic Doppler factor $\delta$=1/$\Gamma$(1-$\beta$cos$\theta$), that relates the intrinsic and observed flux for a source moving at relativistic speed (\cite[Urry \& Padovani 1995]{UrryPadovani95}). The flux enhancement is strongly dependent on the viewing angle and decreases very rapidly for $\theta >$8-10$^{\circ}$.
This trend was confirmed by the detection of only three possible $\gamma$-ray counterparts of RGs by the EGRET telescope onboard the CGRO satellite, i.e. NGC~6251, Centaurus~A and 3C~111 (\cite[Nolan et al. 1996]{nolan96}, \cite[Mukherjee et al. 2002]{reshmi02}, \cite[Sguera et al. 2005]{sguera05}, \cite[Hartmann et al. 2008]{hartmann08}).
A huge step forward in this field has been made thanks to the \emph{Fermi} satellite, as predicted by several works suggesting these sources as possible targets detectable by the Large Area Telescope (LAT) instrument onboard  \emph{Fermi} (\cite[Stawarz et al. 2003, 2006]{stawarz03}, \cite[Ghisellini et al. 2005]{ghisellini05}, \cite[Grandi \& Palumbo 2007]{grandipalumbo07}).

In the following, the main recent results on RGs obtained with the \emph{Fermi}-LAT telescope in the MeV-GeV band (Sec.~2) and with the current Imaging Atmospheric Cherenkov Telescopes in the TeV regime (Sec.~3) will be presented. Within this context, the impact of the future Cherenkov Telescope Array (CTA) on the study of RGs at very-high energies will be also discussed. 

\section{The MeV-GeV band: Fermi-LAT results}
\noindent
The {\it Fermi} $\gamma$-ray space telescope (\cite[Atwood et al. 2009]{atwood09}) was launched in 2008 June 11th and carries onboard two instruments: (i) the LAT, that operates in the nominal energy band 20~MeV-300~GeV \footnote{With the new Pass8 data the LAT energy range extends up to 1~TeV (\cite[Atwood et al. 2013]{atwoodpass8}). For more information about Pass8 \ttfamily{https://fermi.gsfc.nasa.gov/ssc/data/analysis/documentation/Pass8$\_$usage.html}} and (ii) the Gamma-ray Burst Monitor (GBM), that covers the energy range 8~keV-40~MeV. Results presented throughout the paper are based on LAT data.\\

\noindent{\bf Radio galaxies are $\gamma$-ray emitters}

\begin{figure}[h]
\begin{center}
 \includegraphics[width=4.8in]{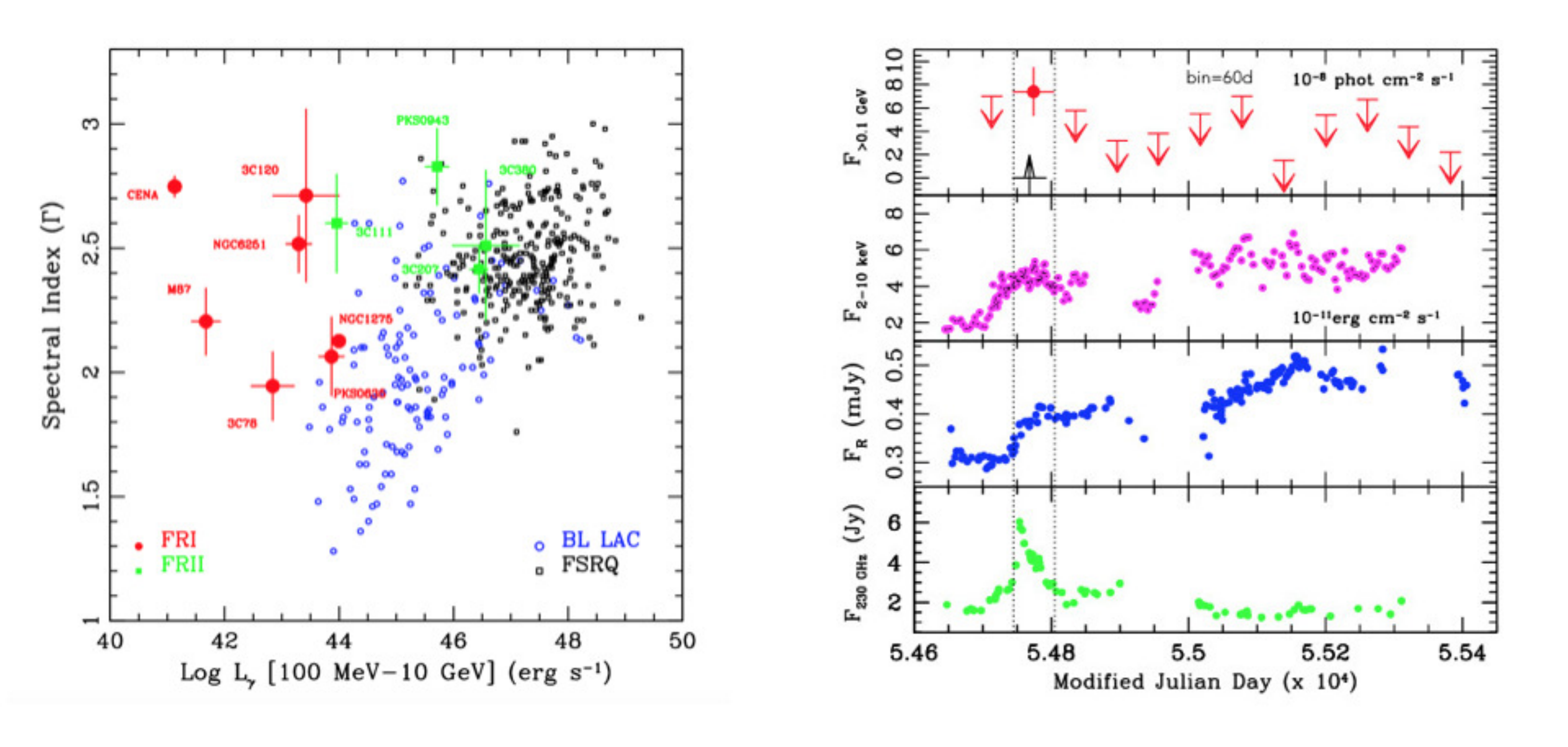} 
 \caption{{\it Left panel:} Power-law photon index ($\Gamma$) plotted as a function of the 100~MeV-10~GeV 
 $\gamma$-ray luminosity for: BL Lacs (open blue circles), FSRQ (open black squares), FR~I (red circles), FR~II (green squares). Blazars and MAGN (radio galaxies and SSRQ) occupy different regions of the plot. The two green squares falling in the FSRQ region are the two more distant sources of the sample, i.e. the two SSRQ 3C~207 and 3C~380. {\it Right panel:} mm (230~GHz; {\it green}), optical (R; {\it blue}), X-ray (2.4-10~keV; {\it magenta}) and $\gamma$-ray (0.1-100~GeV; {\it red}) light curves of 3C~111. mm-to-X-ray data are from \protect\cite[Chatterjee et al. (2011)]{chatterjee11}, $\gamma$-ray data are from \protect\cite[Grandi et al. (2012)]{grandi3c111}. The {\it black dotted lines} limit the flare event  in late 2008 (October 4-December 4). The {\it black arrow} marks the time of ejection of the radio knot. {\it Red downward arrows} are $\gamma$-ray 2$\sigma$ upper limits.}
    \label{fig1}
\end{center}
\end{figure}

\noindent
In only 15 months of survey the LAT could detect 11 radio objects, 7 FR~Is and 4 FR~IIs (e.g. 2 radio galaxies and 2 steep spectrum radio quasars\footnote{Steep Spectrum Radio Quasars are usually considered as FR~II broad line radio galaxies at high redshift.}), generally referred to as Misaligned AGN (MAGN \footnote{With the term Misaligned AGN we consider objects characterised by steep radio spectra ($\alpha>$0.5) and possibly showing symmetrical extension in the radio maps, i.e. radio galaxies and steep spectrum radio quasars.}; \cite[Abdo et al. 2010a]{magn}), confirming that RGs are GeV emitters (Table \,\ref{tab1}).
Most of the sources are faint (F$_{> \rm 0.1~GeV} \sim$10$^{-8}$~ph~cm$^{-2}$~s$^{-1}$) and have steep power-law spectra ($\Gamma >$2.4), in agreement with the Unified Models of AGN in which radio galaxies are considered a de-boosted version of blazars (Fig.\,\ref{fig1} \emph{left panel}).
The  number of radio galaxies caught by the LAT has continuously increased, passing from 11  to 22 in 4 years of operations (3 LAC; \cite[Ackermann et al. 2015]{3lac}), with FR~Is preferentially detected by the LAT with respect to FR~IIs (Table \,\ref{tab1}).
The discrepancy in the FR~I/FR~II detection rate does not seem to be related to a larger distance of FR~IIs, generally found at higher redshifts (\cite[Grandi \& Torresi 2012]{grandi12}). If a correlation between the radio core flux and the $\gamma$-ray flux (\cite[Ackermann et al. 2011]{ackermann2011}, \cite[Ghirlanda et al. 2011]{ghirlanda11}, \cite[Lico et al. 2017]{lico17}) is assumed, implicitly implying that the emission in the two bands is mainly from the jet, several FR~IIs are expected to be detected above the LAT sensitivity threshold  (see also \cite[Kataoka et al. 2011]{kataoka11}). This result suggests that the predominance of FR~Is in the GeV sky should be related to other causes, for example intrinsic differences in the jet structure.\\

\begin{table}
  \begin{center}
  \caption{3LAC MAGN: (i) Source name; (ii) 3FGL name; (iii) right ascension; (iv) declination; (v) redshift; (vi) radio classification; (vii) detected TeV counterpart. 3C~120 has been added to the list since it is a confirmed $\gamma$-ray source, and also Tol1326-379 and PKS1718-379 recently associated with 3FGL sources.}
  \label{tab1}
 {\scriptsize
  \begin{tabular}{l l l l c c c}\hline 
{Source}      & {3FGL Name}           & {RA}            & {Dec}             &{z}               &{FR type}      &{TeV source}\\ 
                   &                                 & (J2000)        & (J2000)           &                &                              &\\ 
 \hline
3C~78         &3FGLJ0308.6+0408     &03 08 26.22	&+04 06 39.3     &0.02865    &FR~I                          &no\\
IC310          &3FGLJ0316.6+4119    &03 16 42.97	&+41 19 29.61   &0.019        &FR~I                          &yes\\
NGC1275     &3FGLJ0319.8+4130   &03 19 48.16	&+41 30 42.1    &0.0175       &FR~I                        &yes\\
ForA(lobes)  &3FGLJ0322.5-3721     &03 21 37.75	&-37 12 49.1    &0.00587      &FR~I                        &no\\
4C+39.12    &3FGLJ0334.2+3915   &03 34 18.41	&+39 21 24.4   &0.02059      &FR~I                        &no\\
Pictor~A      &3FGLJ0519.2-4542    &05 19 49.72	&-45 46 43.85   &0.03506     &FR~II                       &no\\
PKS0625-35 &3FGLJ0627.0-3529    &06 27 06.72	&-35 29 15.33   &0.05494     &FR~I                        &yes \\
3C~189       &3FGLJ0758.7+3747   &07 58 28.1	&+37 47 11.8    &0.04284     &FR~I                       &no\\
3C~221       &3FGLJ0934.1+3933  &09 35 06.63	&+39 42 06.7    &--              &FR~II/SSRQ                &no\\
3C~264       &3FGLJ1145.1+1935  &11 45 05.0	&+19 36 22.74  &0.02172   &FR~I                           &yes\\
M87            &3FGLJ1230.9+1224   &2 30 49.42	&+12 23 28.04  &0.00428  &FR~I                           &yes\\
CenA(core)  &3FGLJ1325.4-4301   &13 25 27.61	&-43 01 08.8    &0.0018    &FR~I                           &yes\\
3C~303        &3FGLJ1442.6+5156 &14 43 02.76	&+52 01 37.29  &0.14119  &FR~II                         &no\\
NGC6251     &3FGLJ1630.6+8232  &16 32 31.96	&+82 32 16.39  &0.02        &FR~I                         &no\\
3C~111       &3FGLJ0418.5+3813c &04 18 21.27	&+38 01 35.8   &0.0485    &FR~II                        &no\\
CenB           &3FGLJ1346.6-6027   &13 46 49.04	&-60 24 29.35	&0.01292 &FR~I                     &no\\
TXS0348+013  &3FGLJ0351.1+0128 &03 50 57.36	&+01 31 04.91	&1.12     &FR~II/SSRQ         &no\\
3C~207           &3FGLJ0840.8+1315 &08 40 47.58	&+13 12 23.56	&0.681  &FR~II/SSRQ        &no\\
PKS1203+04    &3FGLJ1205.4+0412 &12 06 19.92	&+04 06 12.04	&0.536  &FR~II/SSRQ        &no\\
3C~275.1        &3FGLJ1244.1+1615 &12 43 57.64	&+16 22 53.39      &--        &FR~II/SSRQ       &no\\
3C~380           &3FGLJ1829.6+4844 &18 29 31.78	&+48 44 46.16	&0.695  &FR~II/SSRQ       &no\\
3C~120        & --                           &04 33 11.1    &+05 21 16            &0.033 &FR~I                    &no\\
Tol1326-379  &3FGL1330.0-3818    &13 29 19.2  &-38 14 18             &0.028  &FR~0                    &no\\ 
PKS1718-649 &3FGLJ1728.0-6446  & 17 23 41.0 &-65 00 37        &0.014         &CSO                   &no\\
\hline	 
\end{tabular}
  }
 \end{center}
\vspace{1mm}
\end{table}

\noindent
{\bf Localisation of the $\gamma$-ray emitting region}\\
\noindent
One challenging aspect of the high-energy study of MAGN, and of blazars in general, is the localisation of the $\gamma$-ray emitting region. Indeed, localising where high-energy photons are dissipated with respect to the black hole has a strong impact on the physical models invoked to explain the $\gamma$-ray radiation. The site of  emission is not unique: it can occur at sub-pc scales, within the broad-line region (BLR), or outside it on pc scales, as attested by several studies performed on blazars (\cite[Abdo et al. 2010b]{abdo10b}, \cite[Jorstad et al. 2010]{jorstad10}, \cite[Agudo et al. 2011a, 2011b]{agudo11a}). MAGN are particularly suitable for this kind of study since they offer the opportunity to observe $\gamma$-ray emission processes from sub-pc up to kpc scales: different emission sites imply different seed photons involved in the Inverse Compton process producing the GeV emission\footnote{In the standard leptonic scenario.} (\cite[Tavecchio et al. 2011]{tavecchio11}). 
In order to establish where the high-energy radiation is emitted a multi-wavelength (MW) approach is necessary, as demonstrated by the successful results obtained for the broad-line radio galaxies 3C~111 and 3C~120 \footnote{3C~111 is a FR~II, while 3C~120 although morphologically classified as FR~I, is characterised by a powerful accretion disk. Interestingly, these are the only AGN for which a disk/jet connection has been established (\cite[Marscher et al. 2002]{marscher02}, \cite[Chatterjee et al. 2011]{chatterjee11}).}: in these two sources the radio-to-$\gamma$-ray study allowed to establish a clear link between intense $\gamma$-ray flares and the expulsion of bright superluminal knots in the radio core. In 3C~111,
a MW outburst occurred in late 2008 (Fig.\,\ref{fig1} \emph{right panel}) was attributed to the ejection of a new radio blob (\cite[Grandi et al. 2012]{grandi3c111}): the emitting region was compact ($\leq$0.1~pc) and the event was localised in the radio core at $\sim$0.3~pc from the black hole.
A similar result was obtained by \cite[Tanaka et al. (2015)]{tanaka15} from a six-year MW monitoring of 3C~120. Also in this case the $\gamma$-ray flare was related to the ejection of a knot from the radio core at sub-pc distance from the black hole. This finding was further confirmed by \cite[Casadio et al. (2015)]{casadio15} who could better constrain the site of the emitting region within $\sim$0.13~pc by exploiting mm-VLBI data. In the examples above the emitting region is located within the BLR, however, in other cases like M87 different dissipation zones have been found, depending on the event producing the outburst\footnote{In M87 a long-term MW campaign and very-high energy data have allowed to identify different emitting regions: the TeV flare occurred in 2005 was associated with an X-ray burst of the knot HST-1, at 0.85$''$ from the nucleus. In 2008, the coincidence of another TeV flare with radio flares confined the VHE photon production region within 10 Schwarzschild radii. Finally, the 2010 TeV flare was not accompanied by an enhancement of the radio flux but by an increase in the \emph{Chandra} X-ray flux 3 days after the flare, making hard to unambiguously interpret the data.} (\cite[Aharonian et al. 2006]{aharonian06}, \cite[Harris et al. 2006]{harris06}, \cite[Acciari et al. 2009]{acciari09}, \cite[Abramowski et al. 2012]{abramowski12}). 
Although the potentiality of such studies is great, MAGN having dedicated long-term MW campaigns are a few. Multi-frequency monitoring is a difficult task that requires a huge effort to collect (quasi-)simultaneous data necessary to find  connections among different wavebands. However, it would be very important to have ongoing MW campaigns on MAGN when the CTA observatory will be operative. This would represent a great step forward in the comprehension of the jet structure and evolution. \\

\noindent
\textit{Extended $\gamma$-ray emission}\\
\noindent
Extended radio lobes represent another possible site of production of high-energy photons (\cite[Cheung 2007]{cheung07}, \cite[Hardcastle et al. 2009]{hardcastle09}) as confirmed by the LAT detection of the lobes in the FR~I radio galaxy Centaurus~A (\cite[Abdo et al. 2010c]{abdocena}).
The detection and imaging of extended $\gamma$-ray emission is challenging given (i) the PSF of the LAT (the 68$\%$ containment PSF is $\sim$0.8~deg at 1~GeV) and (ii) because the emission is isotropic and not boosted by relativistic effects. Indeed, in Centaurus~A it was possible to separate the contribution of the two lobes from the core in 10 months of data thanks to the proximity of the source (z=0.001825) and to the total angular extent of its lobes ($\sim$10~deg).
The flux above 100~MeV of the two lobes accounts for more than half of the total $\gamma$-ray flux of the source and the emission is likely produced through Inverse Compton (IC) scattering of the Cosmic Microwave Background Radiation (CMB).
Very recently, $\gamma$-ray extended emission was also found in the FR~I Fornax~A (\cite[Ackermann et al. 2016]{ackermann16}) in 6.1 years of data. The source is very nearby (z=0.005871) and the radio lobes span $\sim$50$'$, but differently from Cen~A, the core flux accounts for less than 14$\%$ of the total $\gamma$-ray flux. However, an IC/CMB-EBL model (\cite[Georganopoulos et al. 2008]{georga08}, \cite[Dom\'inguez et al. 2011]{dominguez11}) alone is not sufficient to reproduce the extended emission  because it underestimates the high-energy fluxes. A second component, in addition to the leptonic one, is required and this could be of hadronic origin (\cite[McKinley et al. 2015]{mckinley15}). 
Finally, the \emph{Fermi}-LAT detection of a young Compact Symmetric Object (CSO), i.e. PKS~1718-649 (\cite[Migliori et al. 2016]{migliori16}) further supports the non-thermal $\gamma$-ray production in the lobes (for a detailed discussion see \cite[Migliori 2018, these proceedings]{migliori18}). \\

\noindent{\bf FR~I Spectral Energy Distributions}\\
\noindent
Observations of FR~Is at high- (HE) and very-high (VHE) energies\footnote{HE: $>$100~MeV; VHE: $>$100~GeV (see e.g. \cite[Aharonian 2012]{aharonian12}).} have questioned the pure one-zone SSC model, generally applied to reproduce their spectral energy distributions (SED), because it implies bulk velocities of the jet much lower than BL Lacs ($\Gamma \leq$3). This is at odd with the Unified Scenario in which FR~Is are the misaligned counterpart of BL Lacs. The problem is overcome if stratified jets, consisting of different regions moving at different velocities and mutually interacting, are assumed (\cite[Georganopoulos et al. 2003]{georga03}, \cite[Ghisellini et al. 2005]{ghisellini05}, \cite[B\"ottcher et al. 2010]{bottcher10}): these models were successfully applied to the SED of M87 (\cite[Abdo et al. 2009a]{abdo09a}), NGC~6251 (\cite[Migliori et al. 2011]{migliori11}) and NGC~1275 (\cite[Abdo et al. 2009b]{abdo09b}), suggesting that in FR~Is the presence of slower components could play an important role in amplifying the GeV emission. Interestingly, a limb-brightened jet structure has been observed in M87 (\cite[Kovalev et al. 2007]{kovalev07}) and recently in NGC~1275 (\cite[Giovannini et al. 2018]{giovannini18}, Giovannini 2018, these proceedings).
In FR~IIs the external layers could be less prominent or even absent and/or the deceleration processes less efficient, explaining their rarity in the GeV sky. Other possible sources of $\gamma$-ray photons could be magnetic reconnection events along the jets or in the vicinity of the black hole (\cite[Giannios et al. 2010]{giannios10}, \cite[Khiali et al. 2015]{khiali15}). Hadronic models  based on proton-photon interactions have been also developed to provide connections among AGN, ultra-high-energy cosmic rays (UHECR) and neutrinos (\cite[B\"ottcher 2012]{bottcher12}, \cite[Becker \& Biermann 2009]{becker09}).\\

\noindent
\textit{Temporal variability}\\
\noindent
Studying $\gamma$-ray variability is a powerful tool to further explore the jet structure issue in FR~Is and FR~IIs. 
The two classes show a different temporal behaviour:  FR~Is are detected, on average, for most of the time by the LAT and their light curves do not show variability structures, as suggested by the 3FGL variability index\footnote{The variability index is an indicator of the variability of a source on timescales of months. For the 3FGL catalog (\cite[Acero et al. 2015]{3fgl}), an index $>$72.4 indicates that the source is variable at a 99$\%$ confidence level.} TS$_{\rm var} <$72.4 (\cite[Grandi et al. 2013]{grandi13}). On the contrary, FR~IIs are observed only when intense and rapid flares occur (Fig.\,\ref{fig1} \emph{right panel}). Within this picture an outstanding exception is NGC~1275 which shows a blazar-like behavior and a rapid variability on timescales of days to weeks (\cite[Tanada et al. 2018]{tanada18}).\\

\noindent{\bf The FR~0 radio galaxy Tol1326-379}\\
\noindent
The increased exposure time of the LAT survey combined with recent large-area surveys such as SDSS/NVSS/FIRST (\cite[Best \& Heckman 2012]{bh12}) allowed to explore for the first time the GeV emission of RGs at low radio fluxes ($\sim$mJy). This is the realm of FR~0 sources (\cite[Baldi et al. 2015]{baldi15}), i.e. compact radio galaxies (radio size $\leq$5~kpc) that represent the bulk of the RL AGN population in the local Universe (\cite[Baldi et al. 2018]{baldi18}). 
\cite[Grandi et al. (2016)]{grandi16} proposed the association of a 3LAC source with the FR~0 Tol1326-379, offering the opportunity to investigate for the first time the behavior of these objects in the high-energy domain. FR~0s are similar to FR~Is in the optical and X-ray properties, i.e. they are classified as low-excitation galaxies and their X-ray radiation is of non-thermal origin, likely produced by the jet (\cite[Torresi et al. 2018]{torresi18}). 
In $\gamma$-rays, the flux of Tol1326-379 is similar to FR~Is, $F_{(>1~GeV)}=(3.1\pm0.8)\times10^{-10}~\rm ph~\rm cm^{-2}~\rm s^{-1}$ but the spectrum is steeper, $\Gamma$=2.78$\pm$0.14. 
The SED of Tol1326-379  resembles that of Cen~A,  known to lie close to the plane of the sky, $\theta \sim50-80^{\circ}$.
Therefore, also for the compact FR~0 source the orientation angle could be large and the process responsible for the $\gamma$-ray emission is likely the same as extended FR~Is.\\

\section{Radio galaxies in the TeV regime}
\begin{figure}[h]
\begin{center}
 \includegraphics[width=2.2in]{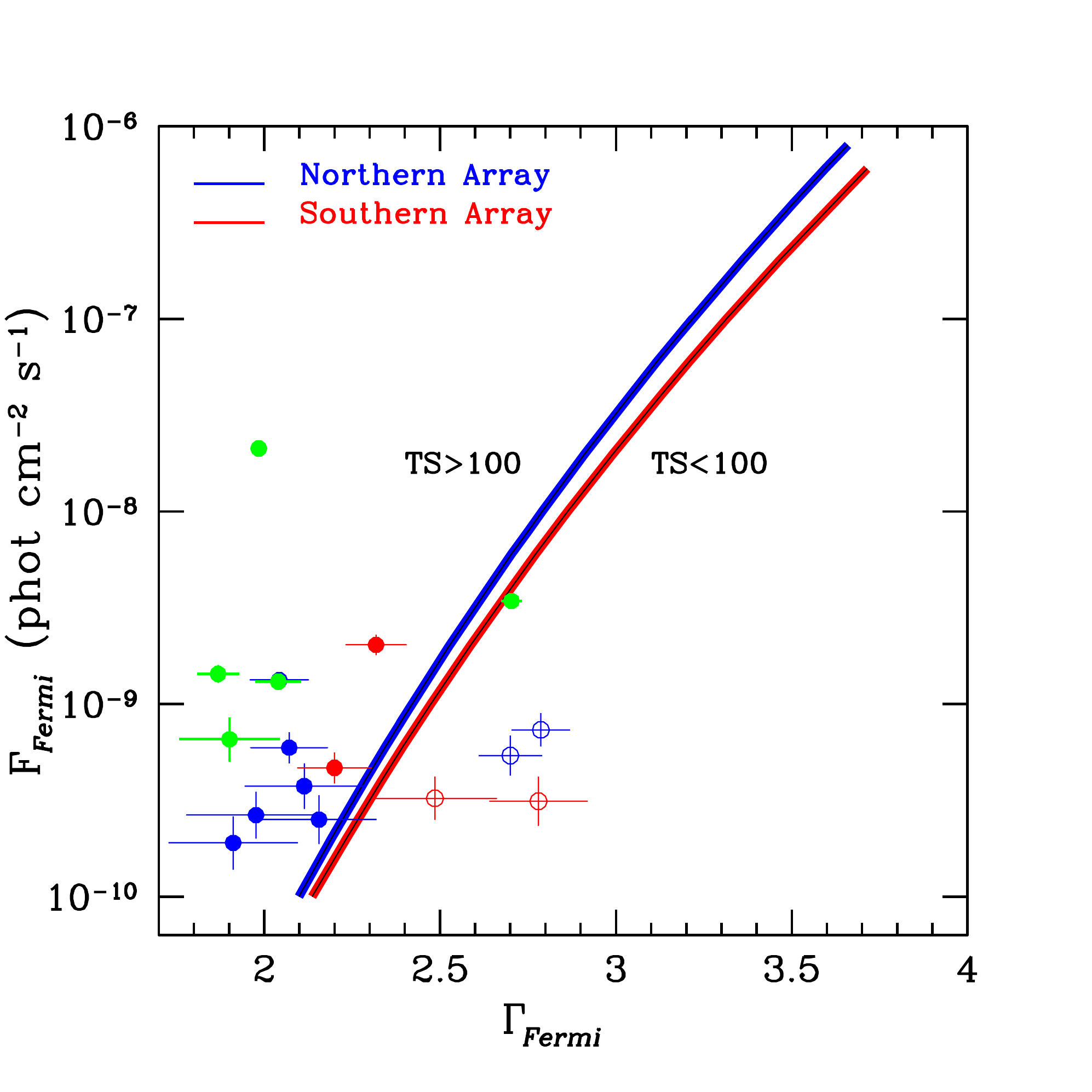} 
 \caption{Diagnostic plot for MAGN with known \emph{Fermi}-LAT flux and spectral slope reproduced from \cite[Angioni et al. (2017)]{angioni17}. The \emph{blue} and \emph{red lines} define the region where a source can be detected by CTA North and South, respectively, assuming a threshold TS$=$100. \emph{Green filled circles} are radio galaxies already detected by current IACT telescopes (before the detection of PKS~0625-35 and 3C~264). \emph{Blue filled circles} are TS$>$100 detections with the northern array and \emph{red filled circles} with the southern array. \emph{Empty circles} represent  TS$<$100 detections.}
   \label{fig3}
\end{center}
\end{figure}

\noindent
Up to now, only a few radio galaxies have a very-high energy counterpart (Table~1): all of them are nearby (z$<$0.06) FR~Is.
Despite their limited number, observations of radio galaxies in the TeV band have already produced intriguing results. For example, (i) the synergy between VHE data and MW observations have allowed to localise the emitting regions in M87  (Sec.~2); (ii) the fast ($\Delta$T$\sim$4.8~minutes) TeV variability, recently caught by MAGIC in IC~310 (\cite[Aleksi\'c et al. 2014]{aleksic14}), has questioned the classical shock-in-jet scenario (\cite[Blandford et al. 1979]{blandford79}) invoking sub-horizon scale structures as a possible explanation for the extreme temporal behavior (Glawion 2018, these proceedings); (iii) the extrapolation of \emph{Fermi}-LAT data to higher energies has evidenced that a second hard component, whose nature is still debated, is necessary to account for the H.E.S.S. flux (\cite[Sahakian et al. 2013]{sahakian13}). It would be important to observe such a structure in other radio galaxies in order to understand if it is (or not) a common characteristic of MAGN.
A major problem at the moment, is the difficulty in extrapolating the model from the GeV to the TeV band since data are often not simultaneous. Hopefully, this issue will be overcome with CTA that working as an observatory in synergy with other facilities, will provide simultaneous data in a very wide energy band from a few MeV up to hundreds TeV.
Thanks to the improvement in sensitivity by a factor of 5x to 20x, depending on the energy range, with respect to the current Cherenkov telescopes, CTA will give a significant contribution in characterising MAGN at very-high energies (\cite[Angioni et al. 2017]{angioni17}). Fig.\,\ref{fig3} is a diagnostic plot showing the detectability of a source by CTA (northern and southern arrays) given a LAT spectral slope and a flux in the 1-100~GeV band: sources with $\Gamma_{\rm LAT}\leq$2.1 can be easily revealed for fluxes down to 10$^{-10}$ph~cm$^{-2}$~s$^{-1}$. As the slope steepens, larger fluxes are required to overcome the sensitivity threshold of the array. Incidentally, we note that this plot predicted the possible detection of PKS~0625-35 and 3C~264, later revealed by H.E.S.S. and VERITAS, respectively (\cite[Dyrda et al. 2015]{dyrda15}, \cite[H.E.S.S. collaboration 2018]{hess18}, \cite[Mukherjee 2018]{mukherjee18}).

\section{Summary and conclusions}
\noindent
In the last decade our view of RGs above 100~MeV has greatly changed.
Indeed, the \emph{Fermi} satellite confirmed RGs as a class of $\gamma$-ray emitters, in spite of the unfavourable inclination angle of their jets. Although they are a small number with respect to the blazars detected by \emph{Fermi} ($<$2$\%$ in the 3 LAC), studying RGs at high- and very-high energies is of great importance to understand the jet phenomenon and to reveal the jet structure complexity. In particular, the observation of RGs in the GeV band has evidenced that: (i) $\gamma$-ray photons can be produced in different sites and at different scales along the jet and in the radio lobes; (ii) a one-zone SSC model used to reproduce the SED of FR~Is is oversimplified: structured jets, magnetic reconnection events or hadronic processes have been invoked to reconcile the data with the Unified Models; (iii) the different detection rate and temporal variability of FR~Is and FR~IIs suggest possible intrinsic jet differences between these two classes.\\
RGs in the TeV band are almost unexplored, however the few detections obtained so far with the current IACTs already produced promising results, such as the sub-horizon scale variability in IC~310 or the second hard component found in Cen~A. CTA will have a strong impact on our comprehension of RGs as a class by detecting additional sources and better characterising those already known to emit TeV $\gamma$-ray photons. This will be very important to shed light on the mechanism producing the VHE radiation, e.g. leptonic or hadronic processes or both. Indeed, RGs are considered good candidates as extragalactic cosmic-ray accelerators, e.g. the relativistic jets and the radio lobes offer an ideal environment where particles can be accelerated, and thus they could be sources of high-energy neutrinos (\cite[Fraija et al. 2016]{fraija16}, \cite[Hooper 2016]{hooper16}, \cite[Murase 2018]{murase18}, \cite[Matthews et al. 2018]{matthews18}).

\section*{Acknowledgments}
\noindent
{\scriptsize 
The CTA Consortium is gratefully acknowledged. ET would like to thank the two reviewers of the CTA SAPO for their useful comments.
ET thanks P.~Grandi, G.~Migliori and G.~Giovannini for reading the manuscript and for insightful suggestions.
The
Fermi
LAT Collaboration acknowledges generous ongoing support from a number
of agencies and institutes that have supported both the development and the operation of the LAT as well as scientific data analysis.  These include the National Aeronautics and Space Administration and the Department of Energy in the United States, the Commissariat `a l'Energie Atomique and the Centre National
de la Recherche Scientifique / Institut National de Physique Nucleaire et
de Physique des Particules in France, the Agenzia Spaziale Italiana and the Istituto Nazionale di Fisica Nucleare in Italy,
the Ministry of Education, Culture, Sports, Science and Technology
(MEXT), High Energy Accelerator Research Organization (KEK)
and Japan Aerospace Exploration Agency (JAXA) in Japan, and
the K. A. Wallenberg Foundation, the Swedish Research Council
and the Swedish National Space Board in Sweden. Additional support for science analysis during the operations phase is gratefully acknowledged from the Istituto Nazionale di Astrofisica in Italy and the Centre National d'Etudes Spatiales in France}

\end{document}